\RequirePackage[mathlines]{lineno} % Display line numbers
\documentclass[prd,twocolumn,showpacs,amsmath,amssymb]{revtex4}
\usepackage{overpic,graphicx}% Include figure files
\usepackage{dcolumn}% Align table columns on decimal point
\usepackage{bm}% bold math
\usepackage{rotating}
\usepackage{subfigure}
\usepackage{color}
\usepackage{multirow}
\usepackage{indentfirst}
\usepackage{amsmath}
\begin{document}
%\linenumbers

\title{\bf \boldmath
Measurements of the Branching Fractions of the Singly-Cabibbo-Suppressed Decays $D^0\to\omega\eta$,
$\eta^{(\prime)}\pi^0$ and $\eta^{(\prime)}\eta$}

\author{
%\begin{small}
%\begin{center}
M.~Ablikim$^{1}$, M.~N.~Achasov$^{9,d}$, S. ~Ahmed$^{14}$, O.~Albayrak$^{5}$, M.~Albrecht$^{4}$, D.~J.~Ambrose$^{46}$, A.~Amoroso$^{51A,51C}$, F.~F.~An$^{1}$, Q.~An$^{48,39}$, J.~Z.~Bai$^{1}$
% ...
\\
\vspace{0.2cm}
(BESIII Collaboration)\\
\vspace{0.2cm} {\it
$^{1}$ Institute of High Energy Physics, Beijing 100049, People's Republic of China\\
% ...
$^{4}$ Bochum Ruhr-University, D-44780 Bochum, Germany\\
$^{5}$ Carnegie Mellon University, Pittsburgh, Pennsylvania 15213, USA\\
% ...
$^{9}$ G.I. Budker Institute of Nuclear Physics SB RAS (BINP), Novosibirsk 630090, Russia\\
% ...
$^{14}$ Helmholtz Institute Mainz, Johann-Joachim-Becher-Weg 45, D-55099 Mainz, Germany\\
% ...
$^{39}$ State Key Laboratory of Particle Detection and Electronics, Beijing 100049, Hefei 230026, People's Republic of China\\
% ...
$^{46}$ University of Rochester, Rochester, New York 14627, USA\\
% ...
$^{48}$ University of Science and Technology of China, Hefei 230026, People's Republic of China\\
% ...
$^{51}$ (A)University of Turin, I-10125, Turin, Italy; (B)University of Eastern Piedmont, I-15121, Alessandria, Italy; (C)INFN, I-10125, Turin, Italy\\
% ...
$^{d}$ Also at the Novosibirsk State University, Novosibirsk, 630090, Russia\\
% ...
%}\end{center}
\vspace{0.4cm}
%\end{small}
}
}

\begin{abstract}
By analyzing a data sample of 2.93 fb$^{-1}$ collected
at $\sqrt s=$ 3.773 GeV with the BESIII detector operated at the BEPCII
storage rings,
we measure the branching fractions
${\mathcal B}(D^0\to\omega\eta)=(2.15\pm0.17_{\rm stat.}\pm0.15_{\rm sys.})\times 10^{-3}$,
${\mathcal B}(D^0\to\eta\pi^0)=(0.58\pm0.05_{\rm stat.}\pm0.05_{\rm sys.})\times 10^{-3}$,
${\mathcal B}(D^0\to\eta^\prime\pi^0)=(0.93\pm0.11_{\rm stat.}\pm0.09_{\rm sys.})\times 10^{-3}$,
${\mathcal B}(D^0\to\eta\eta)=(2.20\pm0.07_{\rm stat.}\pm0.06_{\rm sys.})\times 10^{-3}$ and
${\mathcal B}(D^0\to\eta^\prime\eta)=(0.94\pm0.25_{\rm stat.}\pm0.11_{\rm sys.})\times 10^{-3}$.
We note that ${\mathcal B}(D^0\to \omega\eta)$ is measured for the first time
and that ${\mathcal B}(D^0\to \eta\eta)$ is measured with
much improved precision.
\end{abstract}


\maketitle

\oddsidemargin  -0.2cm
\evensidemargin -0.2cm

\section{Introduction}

Hadronic decays of charmed mesons open a window to explore
the interplay between weak and strong interactions. Based on
flavor SU(3) symmetry, different topological amplitudes for
two-body hadronic decays of $D$ mesons can be extracted by
diagrammatic approach~\cite{bhattacharya,chenghy,prd93} or factorization-assisted topological-amplitude approach~\cite{prd89}.
Consequently, comprehensive measurements of
their branching fractions (BFs) can not only
test the theoretical calculations, but also shed
light on the understanding of SU(3)-flavor
symmetry-breaking effects in $D$ decays~\cite{kwong}.

\begin{figure*}[htbp]
\begin{center}
\subfigure{\includegraphics[width=0.30\textwidth]{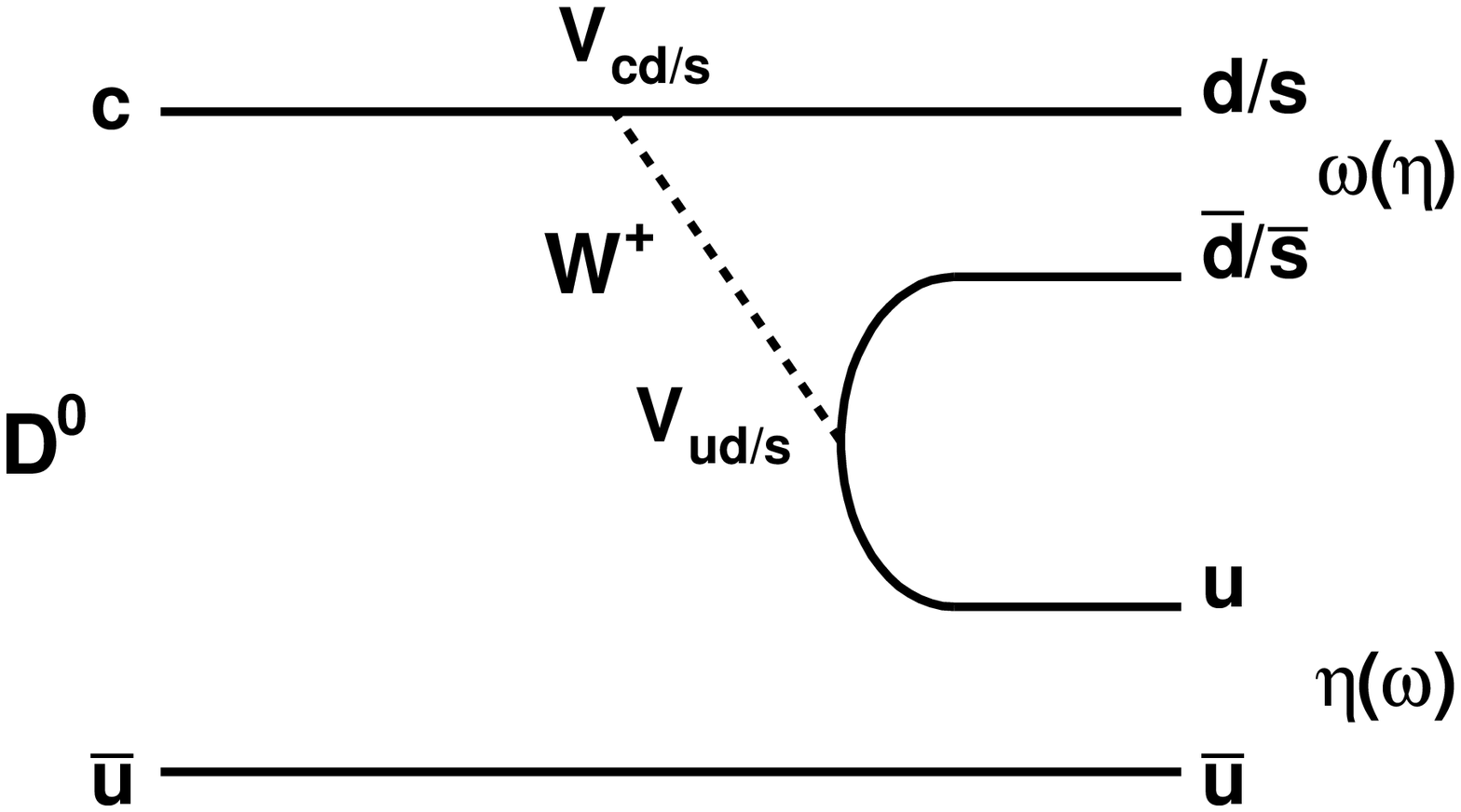}}
\subfigure{\includegraphics[width=0.30\textwidth]{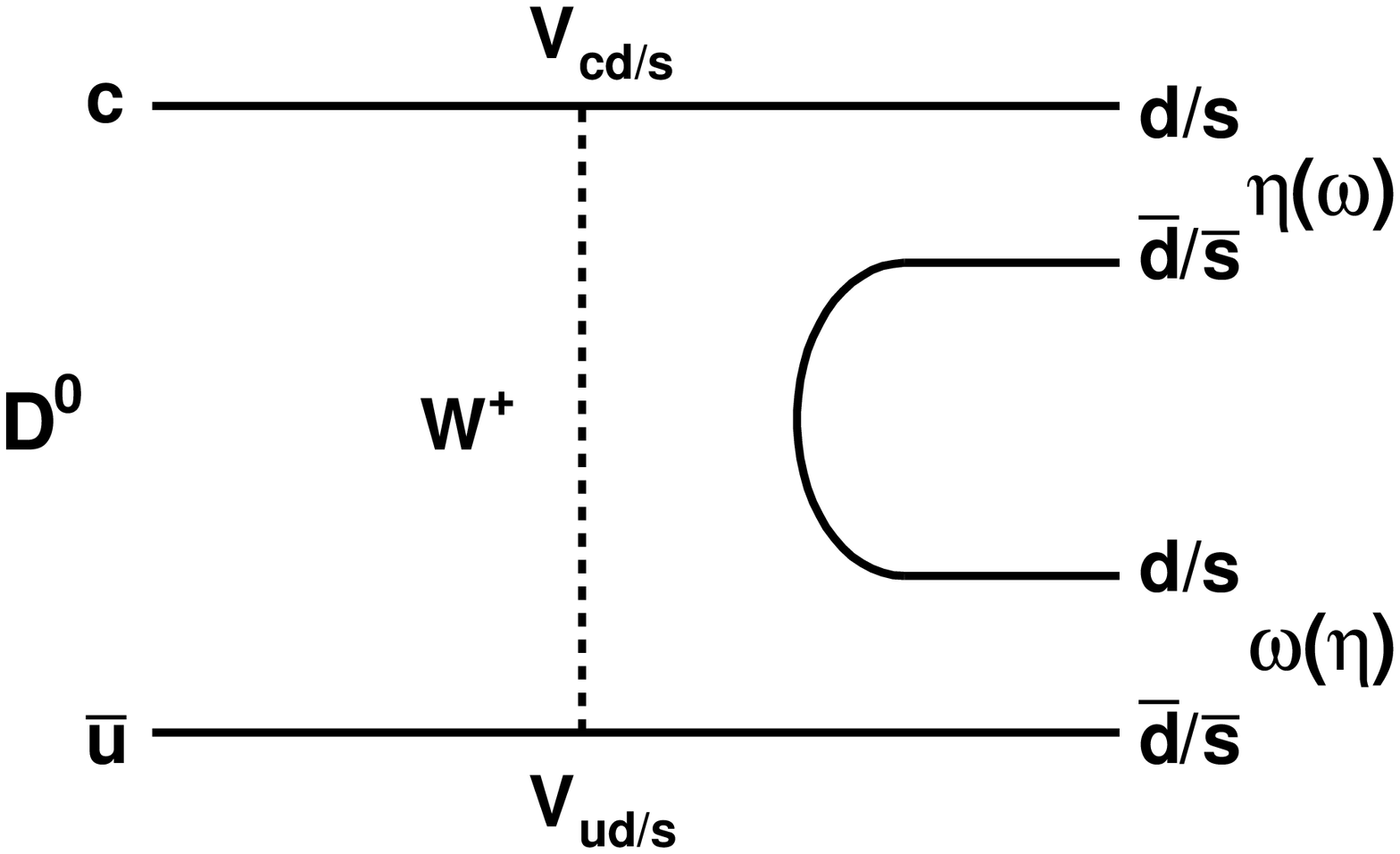}}
\subfigure{\includegraphics[width=0.30\textwidth]{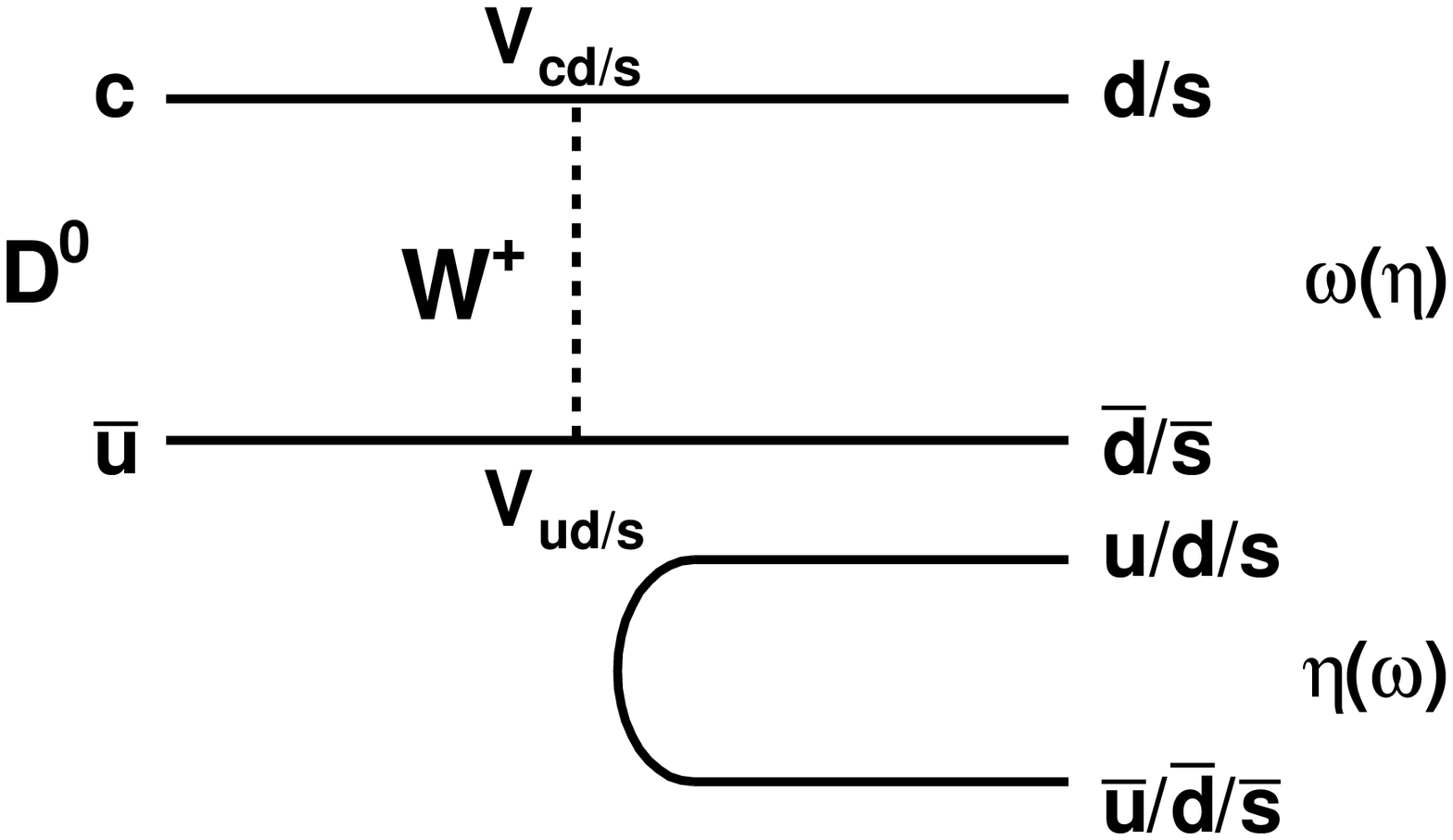}}
\caption{
The Feynman diagrams for the SCS decay $D^0\to \omega \eta$.}
\label{fig:feyman}
\end{center}
\end{figure*}

Two-body $D$ hadronic decays have been extensively investigated
in previous experiments~\cite{pdg2014}.
However, experimental knowledge of some singly-Cabibbo-suppressed (SCS) decays involving four photons,
\emph{e.g.}, $D^0\to\omega\pi^0$, $\omega\eta$, $\pi^0\pi^0$, $\eta\pi^0$, $\eta^\prime\pi^0$,
$\eta\eta$ and $\eta^\prime\eta$,
is still poor due to low statistics and high backgrounds.
The decay $D^0\to\omega \eta$ is particularly interesting, since it only occurs
via $W$-internal emission
and $W$-exchange, as shown in Fig.~\ref{fig:feyman}, and
its decay BF is expected to be at the $10^{-3}$ level~\cite{chenghy}.
However, it has not yet been measured in any experiment.

Previously, the CLEO Collaboration reported the measurements of the BFs of
$D^0\to\eta\pi^0$, $\eta\eta$, $\eta^\prime\pi^0$, $\eta^\prime\eta$~\cite{prd77,prd81}.
During 2010 and 2011, a data sample with an integrated luminosity of 2.93 fb$^{-1}$~\cite{lum} was
collected with the BESIII detector at a center-of-mass energy
$\sqrt s=$ 3.773 GeV. In $e^+e^-$ annihilations at this
energy, $D$ mesons are produced in pairs with no additional particles
and can serve as an ideal
test-bed to systematically study $D$ decays. With this data sample,
the BFs of the two-body hadronic decays $D^0\to\pi^0\pi^0$~\cite{pi0pi0} and
$D^0\to \omega \pi^0, \eta \pi^0$~\cite{omgpi0} have been previously
measured using single-tagged and double-tagged events, respectively,
in which one and two $D$ mesons are reconstructed in each event.
In this paper,
we report the measurements of the BFs for $D^0\to\omega\eta$, $\eta\pi^0$, $\eta^\prime\pi^0$, $\eta\eta$
and $\eta^\prime\eta$, by analyzing single-tag events
using this data sample.
Throughout this paper, the inclusion of charge-conjugate final states
is implied.

\section{BESIII detector and Monte Carlo Simulation}
The BESIII
detector in Beijing, China, is a cylindrical detector with a solid-angle
coverage of 93\% of $4\pi$ that operates at the BEPCII collider consisting
of the following five main components.
A 43-layer main drift chamber (MDC) surrounding the beam pipe
provides precise determinations of charged particle trajectories and
ionization energy losses ($dE/dx$) for charged particle identification (PID).
An array of time-of-flight counters
(TOF) is located outside the MDC and provides additional information
for PID.
A CsI(Tl) electromagnetic calorimeter (EMC) surrounds the TOF and
is used to measure energies of electromagnetic showers.
A solenoidal superconducting magnet outside the EMC
provides a 1 T magnetic field in the central tracking region of the detector.
The iron flux return yoke of the magnet is instrumented
with 1272 m$^2$ of resistive plate muon counters arranged in nine
layers in the barrel and eight layers in the end-caps.
More details of the BESIII detector are described in Ref.~\cite{bes3}.

A GEANT4-based \cite{geant4} Monte Carlo (MC) simulation software
package, which includes the geometrical description of
the detector and its response, is used to determine the detection
efficiency and to estimate the potential backgrounds.
An inclusive MC sample produced at $\sqrt s=3.773$ GeV consists of
$D^0\bar D^0$, $D^+D^-$ and non-$D\bar D$ decays of $\psi(3770)$,
initial-state radiation (ISR) production of $\psi(3686)$ and $J/\psi$,
the $q\bar q$ ($q=u$, $d$,$s$) continuum process,
and Bhabha scattering, di-muon and di-tau events.
The $\psi(3770)$ is generated by the MC generator
KKMC~\cite{kkmc}, in which ISR effects \cite{isr} and final state
radiation (FSR) effects~\cite{photons} are considered.
The known decay modes of $J/\psi$, $\psi(3686)$ and $\psi(3770)$
are generated by using BesEvtGen~\cite{evtgen} with BFs quoted from the PDG~\cite{pdg2010},
and the remaining events are generated with LundCharm~\cite{lundcharm}.
The inclusive MC sample corresponds to about 10 times the equivalent luminosity of data.
To determine reconstruction efficiencies, large exclusive MC samples (`signal MC') of
200\,000 events per decay mode are used.

\section{Data analysis}

The two-body $D$ hadronic decays of interest are selected from combinations of
$\pi^0$, $\eta$, $\omega$ and $\eta^\prime$ mesons reconstructed
using $\pi^0\to \gamma\gamma$, $\eta\to \gamma\gamma$,
$\omega\to\pi^+\pi^-\pi^0$ and $\eta^\prime\to\pi^+\pi^-\eta$ decays, respectively.
The $D^0\to\eta\eta$ decay is also reconstructed
using one $\eta$ undergoing a $\gamma\gamma$ decay
and the other decaying to the $\pi^+\pi^-\pi^0$ final state.
In the following, we use $\eta_{\rm \gamma}$ and $\eta_\pi$ in the decay
$D^0\to\eta\eta$ to denote the decay modes $\eta\to\gamma\gamma$
and $\eta\to\pi^+\pi^-\pi^0$, respectively, but simply use $\eta$
for the other $D^0$ decays with a final-state $\eta$
to represent the decay $\eta\to \gamma\gamma$.

The minimum distance of a charged track to the interaction point (IP) is
required to be within 10 cm along the beam direction and within 1 cm
in the perpendicular plane. The polar angle $\theta$
of a charged track with respect to the positron beam direction
is required satisfy $|\cos\theta|<$ 0.93.
PID is performed by using the $dE/dx$ and TOF measurements to calculate
confidence levels
for pion and kaon hypotheses, $CL_{\pi}$ and $CL_{K}$.
Charged pions are required to satisfy $CL_{\pi}>CL_{K}$.

\begin{figure}[htbp]
\begin{center}
\includegraphics[width=3.2in]{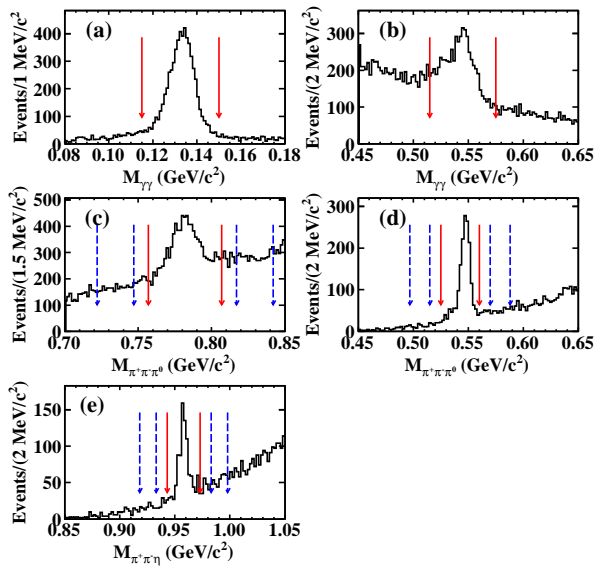}
\caption{
%(Color online) Distributions of the invariant masses for
%(a, b) the $\gamma\gamma$ combinations from the $D^0\to \eta\pi^0$ candidate events,
%(c, d) the $\pi^+\pi^-\pi^0$ combinations from the $D^0\to \omega \eta$ and $D^0\to \eta_{\rm \pi} \eta_{\rm \gamma}$ candidate events,
%(e) the $\pi^+\pi^-\eta$ combinations from the $D^0\to \eta' \pi^0$ candidate events.
%The ranges between the red solid (blue dashed) arrows denote the corresponding signal (sideband) regions.
(Color online) Distributions of the invariant masses for
(a, b) the $\gamma\gamma$ combinations from the $D^0\to \eta\pi^0$ candidate events,
(c, d) the $\pi^+\pi^-\pi^0$ combinations from the $D^0\to \omega \eta$ and $D^0\to \eta_{\rm \pi} \eta_{\rm \gamma}$ candidate events,
(e) the $\pi^+\pi^-\eta$ combinations from the $D^0\to \eta' \pi^0$ candidate events.
The ranges between the red solid (blue dashed) arrows denote the corresponding signal (sideband) regions.
}
\label{fig:signal}
\end{center}
\end{figure}

Photon candidates are chosen from isolated EMC clusters
with energy larger than 25 (50) MeV if the crystal with the maximum
deposited energy in that cluster is in the barrel (end-cap) region~\cite{bes3}.
Clusters due to electronic noise or beam backgrounds are
suppressed by requiring clusters to occur no later than 700 ns
from the event start time.
To reject photons from bremsstrahlung or from secondary interactions,
showers within an angle of $10^\circ$ of the location of charged particles
at the EMC are rejected.
For $\pi^0$ and $\eta_{\rm \gamma}$ reconstruction,
the $\gamma\gamma$ invariant mass is required to be within
$(0.115, 0.150)$ and $(0.515, 0.575)$ GeV/$c^2$, respectively.
To improve $\pi^{0}$ and $\eta_{\rm \gamma}$ momentum resolution,
a kinematic fit is performed to constrain
the $\gamma\gamma$ invariant mass to the appropriate
world average mass~\cite{pdg2014}.
The four-momenta of the $\gamma\gamma$ combinations from the kinematic fit
are used in further analysis.
Since there are two $\eta$ mesons in the final state of the $D^0\to\eta'\eta$ decay,
the $\pi^+\pi^-\eta$ combination with invariant mass closer to
the world average $\eta'$ mass~\cite{pdg2014} is regarded as the $\eta'$ candidate.
Figure~\ref{fig:signal} illustrates the distributions
of the $\gamma\gamma$, $\pi^+\pi^-\pi^0$ and $\pi^+\pi^-\eta$
invariant masses for $\pi^0$ and $\eta_{\rm \gamma}$,
$\omega$ and $\eta_\pi$, and $\eta'$ candidates from data, after above requirements.
In all cases, our nominal $\Delta E$ requirements are applied,
and $M_{\rm BC}$ is required to be in the interval
$(1.860,1.870)$ GeV/$c^2$. See the next paragraph for details about the definitions of $\Delta E$ and $M_{\rm BC}$.
%described below in detail.
%For the fits to the invariant mass spectra of these decay daughters,
%combinatorial backgrounds
%are modeled by a 2nd-order Chebychev polynomial function,
%the
%$\pi^0$, $\eta_{\rm \gamma}$ and $\eta'$ signals are modeled by a Gaussian function,
%while $\omega$ and $\eta_{\rm \pi}$ signals are described by a Breit-Wigner function~\cite{BW}
%and a Crystal-Ball function~\cite{CB}, respectively.
For $\eta_{\rm \pi}$, $\omega$ and $\eta'$ signals,
the $\pi^+\pi^-\pi^0$ and $\pi^+\pi^-\eta$
invariant masses are required to be within signal regions as shown in
Table~\ref{tab:sideband}.

\begin{figure}[htbp]
\begin{center}
\includegraphics[width=2.8in]{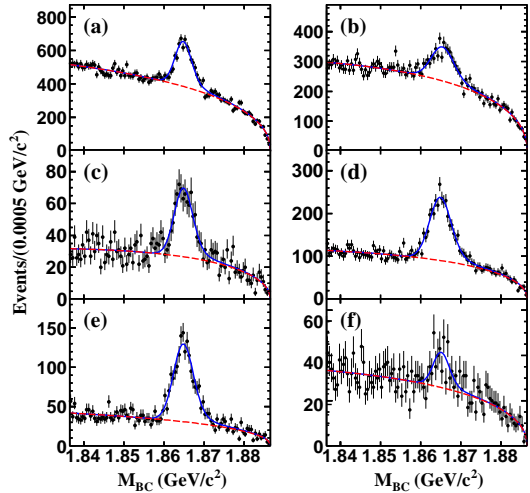}
\caption{
(Color online) Fits to the $M_{\rm BC}$ distributions of the
(a) $D^0\to\omega\eta$,
(b) $D^0\to\eta\pi^0$,
(c) $D^0\to\eta'\pi^0$,
(d) $D^0\to\eta_{\rm \gamma}\eta_{\rm \gamma}$,
(e) $D^0\to\eta_{\rm \pi}\eta_{\rm \gamma}$ and
(f) $D^0\to\eta'\eta$
candidate events in data.
The points with error bars are data.
The blue curves are the total fit results;
the red dashed curves are the background components.
}
\label{fig:fit_mbc1}
\end{center}
\end{figure}

For each selected $D^0$ candidate, two variables, the energy difference
$\Delta E=E_{D^0}-E_{\rm beam}$ and the beam energy constrained mass
$M_{\rm BC} = \sqrt{E^{2}_{\rm beam}/c^4-|\vec{p}_{D^0}|^{2}/c^2}$
are calculated, where $E_{\rm beam}$ is the beam energy, $E_{D^0}$ and
$\vec{p}_{D^0}$ are the energy and momentum of the $D^0$ candidate
in the $e^+e^-$ center-of-mass system.
In the case of a correct $D^0$ candidate, $\Delta E$ and $M_{\rm BC}$
will peak around zero and the nominal $D^0$ mass~\cite{pdg2014}, respectively.
If multiple candidates are found only the combination with the smallest
$|\Delta E|$ is kept in each single-tag mode.
To suppress combinatorial background, mode-dependent $\Delta E$
requirements are
imposed on the
candidates. These correspond approximately to $3\sigma_{\Delta E}$
around the fitted $\Delta E$ peak, where $\sigma_{\Delta E}$ is the
fitted resolution of the $\Delta E$ distribution.
To obtain single-tag $D^0$ yields, we fit the
$M_{\rm BC}$ distributions for each mode,
as shown in Fig.~\ref{fig:fit_mbc1}. In these fits, the
$D^0$ signal is modeled by the MC-simulated shape
convolved with a Gaussian function representing
the mass resolution difference between data and the MC simulation,
and the combinatorial background
is described by an ARGUS function~\cite{ARGUS} with
endpoint fixed to 1.8865 GeV/$c^2$. The parameters of the Gaussian and
ARGUS functions are determined in the fit. The resulting
single-tag $D^0$ yields, $N_{\rm sig}$, are summarized in Table~\ref{tab:bf}.

For the decays containing an $\eta_\pi$, $\omega$ or $\eta'$ meson
in the final state,
the non-$\eta_\pi$, $\omega$ or $\eta'$ contribution in the
$\eta_\pi$, $\omega$ or $\eta'$ signal region is estimated by
using the candidate events within the invariant mass sidebands
listed in Table~\ref{tab:sideband}.
\begin{table*}[htp]
\begin{center}
\caption{\label{tab:sideband}Signal and sideband regions for $\eta_\pi$, $\omega$
and
$\eta'$ mass spectra.}
\begin{tabular}{cccc}
  \hline
  \hline
           & $\eta_{\rm \pi}$ (GeV/$c^2$) & $\omega$ (GeV/$c^2$) & $\eta'$ (GeV/$c^2$)  \\   \hline
Signal region& $(0.525,0.560)$&$(0.757,0.807)$&$(0.943,0.973)$\\
Sideband region& $(0.497,0.515)~or~(0.570,0.587)$&
$(0.722,0.747)~or~(0.817,0.842)$&
$(0.918,0.933)~or~(0.983,0.998)$ \\
  \hline
  \hline
\end{tabular}
\end{center}
\end{table*}
To obtain the single-tag $D^0$ yields in the sideband regions, $N_{\rm sid}$
(see Table~\ref{tab:bf}),
the corresponding $M_{\rm BC}$ distributions are fitted using a method
similar to that described above.
However, due to the low statistics and high backgrounds,
only the parameters of the ARGUS function are left free,
while the parameters of the smearing Gaussian function are fixed to the
values extracted from the $M_{\rm BC}$ fit in the signal region.
The non-$\pi^0$ and non-$\eta_\gamma$ contributions in the $\gamma\gamma$
invariant mass spectra are ignored since decays of the form
$D^0\to \gamma\gamma X$ are highly suppressed, and therefore any combinatoric
background under the $\pi^0$ or $\eta_\gamma$ signals
will not peak in $M_{\rm BC}$.

\section{Results for Branching Fractions}

Detailed MC studies show that, except for the non-resonant
$\eta_\pi$, $\omega$ and $\eta'$ background components,
which are estimated from sideband regions, no other background processes peak in the
$M_{\rm BC}$ distribution.  We may thus determine the BF
for the hadronic decay $D^{0}\to f$ via
\begin{equation}\label{equ:branchingfraction}
{\mathcal B}(D^{0}\to f) = \frac{{N_{\rm net}}}{n\cdot N_{D^{0}\bar{D}^{0}}^{\rm tot}\cdot\epsilon
\cdot {\mathcal B}_{\rm int}}\,.
\end{equation}
Here, $N_{\rm net}$ is the net signal yield,
which is $N_{\rm sig}-N_{\rm sid}$ ($N_{\rm sig}$) when
a sideband subtraction is (is not) applied to the intermediate mass spectra.
The factor $n$ is four for the $D^{0}\to\eta_{\rm \pi}\eta_{\rm \gamma}$
decay and two for other decays.
The common factor of two accounts for charge conjugation, while the
additional factor of two in the $D^0\to \eta_{\rm \pi}\eta_{\rm \gamma}$ decay
accounts for the two possible $\eta_{\rm \pi}\eta_{\rm \gamma}$ combinations
per $D^0$ meson decay.
$N^{\rm tot}_{D^{0}\bar{D}^{0}}$ is the total number of $D^{0}\bar{D}^{0}$ pairs in data,
which is determined to be $(10597\pm28\pm87)\times10^3$~\cite{num},
$\epsilon$ is the detection efficiency,
and ${\mathcal B}_{\rm int}$ denotes the decay BFs
of the intermediate particles $\pi^0$, $\eta_{\rm \gamma(\pi)}$, $\omega$ and $\eta^\prime$~\cite{pdg2014},
which are not included in the detection efficiencies.
The numbers of peaking background events in the
$M_{\rm BC}$ distributions
are assumed to be equal between signal and sideband regions.

The detection efficiencies are estimated by analyzing signal MC events 
with the same procedure as data analysis, and are listed in Table~\ref{tab:bf}.
Detailed studies show that the MC simulated events model data well.

Inserting the numbers of $N_{\rm net}$, $n$,
$N^{\rm tot}_{D^{0}\bar{D}^{0}}$~\cite{num},
$\epsilon$ and ${\mathcal B}_{\rm int}$~\cite{pdg2014} into
Eq.~\eqref{equ:branchingfraction},
we obtain the resultant BFs shown in Table~\ref{tab:bf},
where the uncertainties are statistical only.

\begin{table*}[htp]
\begin{center}
\caption{\label{tab:bf}
Summary of the singly tagged $D^0$ yields ($N_{\rm sig(sid)}$) in the signal (sideband) region in data,
the detection efficiencies ($\epsilon$),
the decay BFs of the
intermediate particles $\pi^0$, $\eta_{(\gamma)(\pi)}$, $\omega$ and $\eta'$~(${\mathcal B}_{\rm int}$)~\cite{pdg2014}, which
are not included in the detection efficiencies and
the measured BFs (${\mathcal B}$).
The uncertainties are statistical only. The symbol `--' denotes that the item is not relevant.}
\begin{tabular}{lccccc} \hline\hline
Decay mode & $N_{\rm sig}$  & $N_{\rm sid}$  & $\epsilon$ (\%) & ${\mathcal B}_{\rm int}$ (\%)& ${\mathcal B}$~$(\times 10^{-3})$    \\
  \hline
$D^0\to\omega\eta$     & $2961\pm146$ & $784\pm97$ & $13.77\pm0.19$ &
34.65& $2.15\pm0.17$\\
$D^0\to\eta\pi^0$                        & $1695\pm144$ & -- & $35.27\pm0.30$
&38.85 & $0.58\pm0.05$\\
$D^0\to\eta'\pi^0$ & $ \hspace{0.15cm}530\pm\hspace{0.15cm}48 $ & $ \hspace{0.15cm}61\pm28$ &
$14.21\pm0.12$ & 8.83 & $0.93\pm0.11$\\
$D^0\to\eta_{\rm \gamma}\eta_{\rm \gamma}$                         & $2123\pm\hspace{0.15cm}87 $ & -- & $29.74\pm0.16$
&15.45& $2.18\pm0.09$\\
$D^0\to\eta_{\rm \pi}\eta_{\rm \gamma}$        & $1315\pm\hspace{0.15cm}54 $ & $\hspace{0.15cm}61\pm29 $ & $15.10\pm0.12$
&17.67& $2.22\pm0.11$\\
$D^0\to\eta^\prime\eta$  & $ \hspace{0.15cm}170\pm\hspace{0.15cm}33 $ & $\hspace{0.15cm}12\pm25  $ & $12.01\pm0.10$
& 6.63 & $0.94\pm0.25$\\
  \hline \hline
\end{tabular}
\end{center}
\end{table*}

\section{Systematic uncertainty}
\label{sec:sys}

Sources of systematic uncertainty in the BF measurements
are summarized in Table~\ref{tab:sys_tot} and discussed below.

\begin{itemize}

\item
\emph {$N_{D^0\bar D^0}^{\rm tot}$}:
The uncertainty of the total number of $D^0\bar D^0$ pairs, 0.9\%~\cite{num},
is considered as a systematic uncertainty for each decay.

\item
\emph {$\pi^\pm$ tracking and PID}:
The $\pi^\pm$ tracking and PID efficiencies are studied by
analyzing double-tagged hadronic $D\bar D$ events.
The systematic uncertainty for the
$\pi^\pm$ tracking and PID efficiencies each are assigned to be 1.0\%
per track.  Tracking and PID systematics are each treated as fully
correlated among themselves, but uncorrelated with each other.

\item
\emph {$\pi^0$ and $\eta_{(\gamma)}$ reconstruction}:
The $\pi^0$ reconstruction efficiency is studied by analyzing
double-tagged hadronic decays $D^0\to K^-\pi^+$ and $K^-\pi^+\pi^+\pi^-$
versus $\bar{D^0}\to K^+\pi^-\pi^0$ and $K_{S}^{0}\pi^0$.
The systematic uncertainties of both the $\pi^0$ reconstruction efficiency
and the $\eta_{(\gamma)}$ reconstruction efficiency are found to be 2.0\%.

\item
\emph {$\omega$, $\eta_{\rm \pi}$ or $\eta'$ signal window}:
The signal mass windows are widened by 2 MeV/$c^2$ for
the $\omega$, $\eta_{\rm \pi}$ or $\eta'$ used in
$D^0\to\omega \eta$, $\eta_{\rm \pi}\eta_{\rm \gamma}$.
$\eta^\prime\pi^0$ or $\eta^\prime\eta$ decays.
We then re-determine the BFs, and the resulting differences, ranging from
0.5\% to 3.3\%, are taken as systematic uncertainties.

\item
\emph {$\Delta E$ requirement}:
Our $\Delta E$ requirements are widened from 3 to 3.5 times the fitted width,
and we re-calculate the BFs.
The resulting differences, ranging from 3.0\% to 8.7\%,
are taken as systematic uncertainties.

\item
\emph {$M_{\rm BC}$ fit}:
The uncertainties associated with the $M_{\rm BC}$ fits are estimated by
comparing the nominal BFs to the measured values with alternative signal
yield fits.
Variations include alternative total fit ranges of $(1.8335,1.8865)$
or $(1.8395,1.8865)$ GeV/$c^2$, alternative endpoints of 1.8863
or 1.8867 GeV/$c^2$ for the ARGUS background function, and changes in the
detailed method used to extract the MC signal shape.
The quadratic sum of changes in the BFs,
ranging from 1.5\% to 5.3\%, are taken as the systematic uncertainties.

\item
\emph {Normalization of the backgrounds in signal/sideband regions (BKG normalization)}:
Our nominal sideband subtraction for peaking backgrounds from non-resonant
combinatorics in the $\omega$, $\eta_\pi$ and $\eta'$ spectra assumes
that the equal area of the sideband and signal regions gives a correct
normalization.
This is investigated by using instead a scale factor obtained from fitting
the corresponding $\pi^+\pi^-\pi^0$ or $\pi^+\pi^0\eta$
invariant mass spectra in data and integrating the background shape.
The relative changes of the BFs, ranging from 0.4\% to 1.1\%
are used as systematic uncertainties.

\item
\emph {Intermediate BFs}:
The uncertainties on the quoted BFs for $\pi^0\to\gamma\gamma$,
$\eta\to\gamma\gamma$, $\omega\to\pi^+\pi^-\pi^0$, $\eta\to\pi^+\pi^-\pi^0$ and
$\eta^\prime\to\pi^+\pi^-\eta$ of 0.03\%, 0.5\%, 0.8\%, 1.2\% and 1.6\%~\cite{pdg2014}, respectively, are propagated as systematic uncertainties.

\item
\emph {MC statistics}:
The uncertainties due to limited MC statistics used in determining
efficiencies, varying from 0.5\% to 1.3\%, are included.

\end{itemize}

All the individual systematic uncertainties are summarized in Table~\ref{tab:sys_tot}.
For the measurements of $D^0\to \eta_{\rm \pi}\eta_{\rm \gamma}$
and $D^0\to \eta_{\rm \gamma}\eta_{\rm \gamma}$,
the systematic uncertainties are classified into common and independent parts,
necessary for the proper combination of these two measurements later.
For each decay, the total systematic uncertainty
is the quadratic sum of the individual ones.

\begin{table*}[hbtp]
\begin{center}
\caption{\label{tab:sys_tot}
Systematic uncertainties (\%) of the measured BFs, where
{\it com} and {\it ind} denote the common and independent systematic uncertainties
in the measured BFs for $D^0\to\eta_{\rm \gamma}\eta_{\rm \gamma}$ and $D^0\to\eta_{\rm \pi}\eta_{\rm \gamma}$; the symbol `--' denotes that the uncertainty is not relevant.}
\begin{tabular}{cccccccccc} \hline \hline
\multirow{2}{*}{Source}& $D^0\to\omega\eta$     &
$D^0\to\eta\pi^0$   &$D^0\to\eta^\prime\pi^0$&
\multicolumn{2}{c} {$D^0\to\eta_{\rm \gamma}\eta_{\rm \gamma}$}   &
\multicolumn{2}{c} {$D^0\to\eta_{\rm \pi}\eta_{\rm \gamma}$}  &
$D^0\to\eta^\prime\eta$ \\
& & & &{\it com} &{\it ind} &{\it com} &{\it ind} &  \\ \hline

$N_{D^0\bar D^0}^{\rm tot}$  &0.9&0.9&0.9&0.9&-- &0.9&-- &0.9\\
$\pi^\pm$ tracking                      &2.0&-- &2.0&-- &-- &-- &2.0&2.0\\
$\pi^\pm$ PID                           &2.0&-- &2.0&-- &-- &-- &2.0&2.0\\
$\pi^0$ and $\eta_{(\rm \gamma)}$ reconstruction  &4.0&4.0&4.0&4.0&-- &4.0&-- &4.0\\
$\omega$, $\eta_{\rm \pi}$ or $\eta'$ signal window &0.5 &--  &3.3 &-- &-- &-- &0.9 &1.1\\
$\Delta E$ requirement                &3.9&4.8&7.5&-- &3.1&-- &3.0&8.7\\
$M_{\rm BC}$ fit                      &2.3&5.3&2.5&-- &1.5&-- &1.7&4.5\\
BKG normalization        &0.5&-- &1.1&-- &-- &-- &0.4&0.9\\
Quoted BF                             &0.9&0.5&1.7&0.5&0.5&0.5&1.2&1.7\\
MC statistics                         &1.3&0.8&0.9&-- &0.5&-- &0.8&0.8\\
\hline
Total                               &6.9&8.3&9.6&
\multicolumn{2}{c} { 5.4 } &
\multicolumn{2}{c} { 6.3  }&
11.2 \\
\hline \hline
\end{tabular}
\end{center}
\end{table*}

\section{Summary}

Based on an analysis of the singly tagged events using the data
sample of 2.93 fb$^{-1}$ taken at $\sqrt s=$
3.773 GeV with the BESIII detector,
the BFs of the SCS decays
$D^0\to\omega \eta$, $\eta\pi^0$, $\eta'\pi^0$, $\eta\eta$ and $\eta'\eta$ are measured,
and are summarized in Table~\ref{tab:comparison}.
Here, the first and second uncertainties are statistical and systematic, respectively.
The presented ${\mathcal B}(D^0\to\eta\eta)$ is the combination of two individual measurements,
 ${\mathcal B}(D^0\to\eta_{\rm \gamma}\eta_{\rm \gamma})=(2.18\pm0.09\pm0.12)\times 10^{-3}$ and ${\mathcal B}(D^0\to\eta_{\rm \pi}\eta_{\rm \gamma})=(2.22\pm0.11\pm0.14)\times 10^{-3}$, by using the least squares method~\cite{method}
and incorporating the common and independent uncertainties
between the two modes as shown in Table~\ref{tab:sys_tot}.

We compare the measured BFs and the world-average values, as shown in Table~\ref{tab:comparison}.
The ${\mathcal B}(D^0\to \omega\eta)$ is measured for the first time and its magnitude
is consistent with the theoretical prediction~\cite{chenghy,prd89,prd93}, while
the other four BFs are consistent with the world averaged values within uncertainties,
and are of comparable or significantly improved ($D^0\to \eta\eta$) precision.
These measurements provide helpful experimental data to improve our understanding of
SU(3)-flavor symmetry breaking effects in
$D$ decays~\cite{kwong}.

\begin{table}[hbtp]
\begin{center}
\caption{\label{tab:comparison}
Comparisons of the BFs ($\times 10^{-3}$) measured in this work and
the world averaged values.}
\begin{tabular}{l c c} \hline \hline
Decay mode  & This work & PDG~\cite{pdg2014} \\
\hline
$D^0\to\omega\eta$      &  $2.15\pm0.17\pm0.15$ &  --          \\
$D^0\to\eta\pi^0$       &  $0.58\pm0.05\pm0.05$ & $0.68\pm0.07$\\
$D^0\to\eta^\prime\pi^0$&  $0.93\pm0.11\pm0.09$ & $0.90\pm0.14$  \\
$D^0\to\eta\eta$        &  $2.20\pm0.07\pm0.06$        & $1.67\pm0.20$\\
$D^0\to\eta^\prime\eta$ &  $0.94\pm0.25\pm0.11$ & $1.05\pm0.26$\\
\hline \hline
\end{tabular}
\end{center}
\end{table}

\section{Acknowledgements}
The BESIII collaboration thanks the staff of BEPCII and the IHEP computing center for their strong support. This work is supported in part by National Key Basic Research Program of China under Contract No. 2015CB856700; National Natural Science Foundation of China (NSFC) under Contracts Nos. 11235011, 11305180, 11775230, 11335008, 11425524, 11625523, 11635010; the Chinese Academy of Sciences (CAS) Large-Scale Scientific Facility Program; the CAS Center for Excellence in Particle Physics (CCEPP); Joint Large-Scale Scientific Facility Funds of the NSFC and CAS under Contracts Nos. U1332201, U1532257, U1532258; CAS under Contracts Nos. KJCX2-YW-N29, KJCX2-YW-N45, QYZDJ-SSW-SLH003; 100 Talents Program of CAS; National 1000 Talents Program of China; INPAC and Shanghai Key Laboratory for Particle Physics and Cosmology; German Research Foundation DFG under Contracts Nos. Collaborative Research Center CRC 1044, FOR 2359; Istituto Nazionale di Fisica Nucleare, Italy; Joint Large-Scale Scientific Facility Funds of the NSFC and CAS; Koninklijke Nederlandse Akademie van Wetenschappen (KNAW) under Contract No. 530-4CDP03; Ministry of Development of Turkey under Contract No. DPT2006K-120470; National Natural Science Foundation of China (NSFC) under Contract No. 11505010; National Science and Technology fund; The Swedish Research Council; U. S. Department of Energy under Contracts Nos. DE-FG02-05ER41374, DE-SC-0010118, DE-SC-0010504, DE-SC-0012069; University of Groningen (RuG) and the Helmholtzzentrum fuer Schwerionenforschung GmbH (GSI), Darmstadt; WCU Program of National Research Foundation of Korea under Contract No. R32-2008-000-10155-0.

\end{document}